\documentclass[12pt]{article}
\pdfoutput=1
\usepackage{epsfig}
\usepackage{graphicx}
\usepackage{amsmath}
\usepackage{amscd}
\usepackage{amsfonts}
\usepackage{amssymb}

\newcommand {\btheta}{\mbox{\boldmath $\theta$}}
\newcommand {\bTheta}{\mbox{\boldmath $\Theta$}}

\newcommand {\bmu}{\mbox{\boldmath $\mu$}}

\newcommand {\bLambda}{\mbox{\boldmath $\Lambda$}}

\newcommand {\by}{\textbf{y}}
\newcommand {\bY}{\textbf{Y}}

\newcommand {\bC}{\textbf{C}}
\newcommand {\bG}{\textbf{G}}

\newcommand {\bZ}{\textbf{Z}}

\newcommand {\bS}{\textbf{S}}

\begin{document}

\begin{center}
{\bf\Large Fast Bayesian Semiparametric Curve-Fitting and Clustering in Massive Data With Application
to Cosmology}
\end{center}

\begin{center}
\textsc{
\footnote{S. Mukhopadhyay is a PhD student in Bayesian and Interdisciplinary Research Unit, Indian Statistical Institute, 203, B. T. Road, Kolkata 700108,
S. Roy is a Professor in Physics and Applied Mathematics Unit, Indian Statistical Institute, 203, B. T. Road, Kolkata 700108, and S. Bhattacharya is an Assistant Professor in
Bayesian and Interdisciplinary Research Unit, Indian Statistical Institute, 203, B. T. Road, Kolkata 700108.
Corresponding e-mail:
{\it sabyasachi\_r@isical.ac.in}.} Sabyasachi Mukhopadhyay, Sisir Roy, and Sourabh Bhattacharya}
\end{center}

\begin{abstract}

Recent technological advances have led to a flood of new data on
cosmology rich in information about the formation and evolution of
the universe, {\it e.g., the data collected in Sloan Digital Sky
Survey (SDSS) for more than 200 million objects}. The analyses of
such data demand cutting edge statistical technologies. Here, we
have used the concept of mixture model within Bayesian
semiparametric methodology to fit the regression curve with the
bivariate data for the apparent magnitude and redshift for Quasars
in SDSS (2007) catalogue. Associated with the mixture modeling is
a highly efficient curve-fitting procedure, which is central to
the application considered in this paper. Moreover, we adopt a new
method for analysing the posterior distribution of clusterings,
also generated as a by-product of our methodology. The results of
our analysis of the cosmological data clearly indicate the
existence of four change points on the regression curve and
posssibiltiy of {\it clustering} of quasars specially at high
redshift. This sheds new light not only on the issue of evolution,
existence of acceleration or decceleration and environment around
quasars at high redshift but also help us to estimate the
cosmological parameters related to acceleration or decceleration.
\\[2mm]

\end{abstract}

\section{Introduction}
\label{sec:intro}
Among many different ways of testing models of cosmological sources, especially quasars, one is
through the investigations of the distributions, ranges and more importantly the correlations among
the relevant physical characteristics, such as luminosity, spectra, redshifts or cosmological distances.
The impossibility of direct measurements to quasi stellar objects prevents one to validate any direct
relationship between distance and redshift where, the measurable quantities are the apparent
magnitude $(m)$, redshift $(z)$ which, as related to luminosity function or, even the probability
distribution of absolute magnitude $(M)$, as predicted by a given cosmology and the angular diameter
of the object. The recent increase of the computing power led the theorists to simulate the realistic
 physical situations in specific details and predictions which are beyond the present limit of
 experimental techniques.
For example, in the case of n-body simulations, it is possible to
answer {\it given certain initial conditions at some time $t_1$
and some assumed laws of physics, what will be the state of the
system at later time $t_2$} Hockney(1988) ?
However, it is harder to solve the {\it inverse
problem} Aster(2004): given all of the data, what can be said about
the laws of physics that have been operating Brewer(2008)?
Various cosmological
models are considered to understand the formation and evolution of
the universe. In cosmology, for a given set of data, there exists
many possible explanations. A typical observation may rule out
some theories but may be consistent with some others. Again, many
specific techniques have been constructed to tackle each inverse
problem seperately. It is worth mentioning that Efron(1992),
Efron(1999) considered different types of statistical arguments
and tests on the truncated data gathered by astronomers to extract
important statistical characteristics. One of the present
authors (SR) along with his collaborators
Roy(2007) used the
non-parametric methods of Efron(1992) and Efron(1999) to study the bivariate distribution of two physically important
parameters i.e. redshift and apparent magnitude observed in SDSS quasar survey (2005). The data is truncated
in nature. However,
the data in SDSS quasar survey of 2007 is no longer truncated. Here, we will discuss a general
framework using concepts of Bayesian mixture models and Dirichlet process to study the existence of
clusterings in the quasar sample for the whole range of redshifts
and the changepoints associated to the non-linearity of the
fitting curve. The existence of non-linearity can be associated
with the certain physical factors like evolution and presence of
different environments around quasars at high redshift. This will
shed new light on the present cosmological debates regarding the
concordant redshift, age of the universe and
acceleration/deceleration parameters.

On the statistical methodological side, we adopt a very fast and efficient
method for learning about posteriors associated with Bayesian mixture models
with unknown number of components. In particular, we adopt a semiparametric Bayesian curve-fitting
procedure based on our mixture model. We demonstrate that our methods are particularly suitable
for application in massive data, as our present cosmological data.
We also adopt a methodology for analysing the posterior distribution of the clusterings associated
with our Bayesian mixture model with unknown number of components. Our methods are broadly based
on the works of Bhattacharya(2008) and Bhattacharya et al(2008); however, important extensions to modeling multivariate
data are described here. Perhaps more importantly, we demonstrate in this paper that cutting-edge
research works of great scientific importance are possible with our methodologies, despite the enormity
of the size of the data sets. To our knowledge, such advancement in the
Bayesian semiparametric/clustering paradigm has not been possible before.

The rest of the paper is structured as follows.
In Section \ref{sec:overview} we explain the data and in this connection, provide a brief overview
of Bayesian mixture models with unknown number of components.
Our Bayesian semiparametric curve fitting method in (massive) data sets
consisting of multivariate observations is introduced in Section \ref{sec:mult_extension}.
A Gibbs sampling algorithm to simulate from the associated posterior distributions is derived
in Section \ref{sec:gibbs_sampling}.
In Section \ref{sec:simstudy_curve} we illustrate our curve-fitting methodology with a simulated data set, and
application to the real cosmological data is considered in Section \ref{sec:real_application}.
Discussion on summarization of the posterior distribution of clusterings is provided in Section \ref{sec:clustering},
and application of the clustering ideas to the real cosmological
data set is considered in Section \ref{sec:clustering_application}.
The implications of our analysis of the cosmological data set, and related future work, are enlisted in
Section \ref{sec:implications}.

\section{The data and overview of mixture models}
\label{sec:overview}

Our massive cosmological data set, consists of 96307 data points on logarithm of redshift ($z$) and apparent magnitude ($m$)
for Qsasars (qsasi-stellar objects) collected from SDSS data.
The data set does not reveal any clear-cut parametric relationship between the two variables of interest; moreover, our exploratory analyses
clearly indicated that the (bivariate) normality assumption does not hold for the data. Indeed, our quantile-quantile plots of each of
the two variables showed that the marginal distributions of both the variables are far from univariate normal.

To resolve this problem we will use idea of mixture models, which are noted for their flexibility. Indeed,
as noted by Dalal(1983) and Diaconis(1985), mixture models composed of standard densities can, in principle, approximate any underlying distribution.
For more on mixture models, see Titterington(1985), McLachlan(1988).
However, a technical problem associated with classical analysis of mixture models is associated with the number of mixture components
included in the model. In the classical statistical literature there does not seem to exist any rigorously procedure of selecting
an adequate number of mixture components. On the other hand, the Bayesian paradigm offers elegant solutions to this problem. Among the
contributions of Bayesians in this topic, notable are those of Escobar(1995) (henceforth, EW)
and Richardson(1997) (henceforth, RG). The former use Dirichlet process to indirectly induce (random) variability in the number of components, while
the approach of RG directly acknowledges uncertainty about the number of components and puts a prior distribution
on the same, thus rendering the problem variable-dimensional. The methodology of RG
relies on reversible jump Markov chain Monte Carlo (RJMCMC) Green(1995) for drawing inference.

However, it is important to note that the RJMCMC method proposed by RG is quite complicated, and is error prone.
But of more concern is the fact that their methodology is extremely sensitive to the ``move types" selected, and since there are
no general guidelines for selecting optimum move types, the algorithm could be very inefficient. Moreover, for variable-dimensional
problems diagnosis of convergence of RJMCMC is a serious problem. The aforementioned problems asociated with the RJMCMC method
are of course many times aggravated for multivariate observations. The methodology of EW is not a variable dimensional problem
and straightforward Gibbs sampling methods are available,
however, the number of parameters increases with data size, making Gibbs sampling (or any other sampling methods) infeasible for massive
data sets. In response to this computational challenge Wang(2008) have proposed the sequential updating and greedy search (SUGS) algorithm
which proceeds by cyling through the data points, sequentially allocating them to the cluster that maximizes the conditional posterior
allocation probability. The conditional distribution of the unknown parameter, which admits a closed form expression given the maximizing cluster, is then updated.
A complete sweep of the algorithm yields the conditional posterior distribution of all the parameters, given the seuqentially optimal clusterings.
The advantage of the method of Wang(2008) is that it is quite fast, since it does not
rely upon MCMC methods. The disadvantages are that the method does not have a theoretical basis, in that the correct joint or marginal posterior distributions
of the parameters or clusterings are not obtained. Moreover, although the algorithm produces a sequentially optimal clustering, it does not yield
a global maximum {\it a posteriori} (MAP) estimate. The algorithm depends upon the the order in which we consider the observations.
In case of large data set this problem is tackled by considering a few random ordering of the observations and then using pseudo-marginal likelihood (PML), which makes this method an ad hoc one. Perhaps more critically, the algorithm does not assist in any way in obtaining and studying
the probability distribution of the clusterings.

We avoid all the difficulties noted above by adopting a model which may be viewed as a reconciliation of the methods of EW and RG.
The details are outlined next.

\section{Direct Bayesian mixture modeling of multivariate observations using Dirichlet process and associated Bayesian curve fitting}
\label{sec:mult_extension}

The cosmological data set of our interest is, as already noted above, consists of bivariate observations. As a result,
the model and the methodologies proposed by Bhattacharya(2008) warrants extension to bivariate, in fact, more generally,
to multivariate situations. For the sake of full generality, we extend the proposals of Bhattacharya(2008) to the case of $d$-dimensional
observations, where $d\geq 1$.

We assume for $i=1,\ldots,n$, data set $\bY=\left\{\by_1,\ldots,\by_n\right\}$ is available, where observation $\by_i$ can be modeled as a mixture of $d$-variate normal distributions, having $p$ components.
Crucially, $p$ is assumed to be unknown. Rather than assuming a prior distribution on $p$ like RG and treating the problem as variable dimensional,
we assume the following form of mixture representation of the $d$-variate observation $\by_i$:
\begin{equation}
[\by_i\mid\bTheta_M]=\frac{1}{M}\sum_{j=1}^M\frac{|\bLambda_j|^{\frac{1}{2}}}{(2\pi)^{\frac{d}{2}}}\exp\left\{-\frac{1}{2}\left(\by_i-\bmu_j\right)'\bLambda_j\left(\by_i-\bmu_j\right)\right\}
\label{eq:mult_normixture}
\end{equation}
In the above, $M (\geq p)$ is the maximum number of components the mixture can possibly have, and is known;
$\bTheta_M=\{\btheta_1,\ldots,\btheta_M\}$, with $\btheta_j=(\bmu_j,\bLambda_j)$.
We further assume that $\bTheta_M$ are samples drawn from a Dirichlet process (see, for example, Ferguson(1973), EW)\\
\begin{center}
$\btheta_j$ is  iid  from \bG\nonumber\\
{\bG}  is  from  $DP(\alpha\bG_0)\nonumber$
\end{center}

A crucial feature of our modelling style concerns the discreteness
of the prior distribution $\bG$, given the assumption of Dirichlet process; that is, under these assumptions,
the parameters $\btheta_{\ell}$ are coincident with positive probability.
In fact, this is the property that can be exploited to show that (\ref{eq:mult_normixture}) boils down to the form
\begin{equation}
[\by_i\mid\bTheta_M]=\sum_{j=1}^p\pi_j\frac{|\bLambda^*_j|^{\frac{1}{2}}}{(2\pi)^{\frac{d}{2}}}\exp\left\{-\frac{1}{2}\left(\by_i-\bmu^*_j\right)'\bLambda^*_j\left(\by_i-\bmu^*_j\right)\right\}
\label{eq:mult_normixture2}
\end{equation}
where $\left\{\btheta^*_1,\ldots,\btheta^*_p\right\}$ are $p$ distinct components in $\bTheta_M$ with $\btheta^*_j$ occuring $M_j$ times,
and $\pi_j=M_j/M$. Hence, although our model is actually variable dimensional, this is induced through the Dirichlet process prior,
and does not involve complexities as in RJMCMC. In fact, we will derive an easily implementable Gibbs sampling algorithm, even for
highly multivariate observaions.
Observe that, in sharp contrast to the proposed model of EW, the number of parameters to be simulated remains fixed (since
the maximum number of mixture components is fixed), even though the number of observations, $n$, could be extremely large.

Associated with the mixture model (\ref{eq:mult_normixture}) is the idea of Bayesian curve-fitting.
This we illustrate in the next section.

\subsection{Bayesian curve fitting}
\label{subsec:curve_fitting}

In simplified notation, we write (\ref{eq:mult_normixture}) as
\begin{equation}
[\by\mid\bTheta_M]=\frac{1}{M}\sum_{j=1}^MN_d\left(\by:\bmu_j,\bLambda^{-1}_j\right)
\label{eq:mult_normixture3}
\end{equation}
It follows that the conditional distribution of $y_{1}$ given $\by_{-1}=(y_{2},\ldots,y_{d})'$ is
given by
\begin{equation}
[y_{1}\mid\bTheta_M,\by_{-1}]=
\frac{1}{M}\sum_{j=1}^MN_{d-1}\left(\by_{-1}:\bmu_{-1j},\bLambda^{-1}_{-1j}\right)
\times
N\left(y_{1}:\mu^{(j)}_{1|2,\ldots,d},\lambda^{(j)}_{1|2,\ldots,d}\right)
\label{eq:mult_normixture4}
\end{equation}
where $\mu^{(j)}_{1|2,\ldots,d}$ and $\lambda^{(j)}_{1|2,\ldots,d}$ are, respectively,
the univariate conditional mean
$E(y_{1}\mid\by_{-1},\bTheta_M)$ and the inverse precision $1/V(y_{1}\mid\by_{-1},\bTheta_M)$ under the assumption
that {\by} is from $N_d(\bmu_j,\bLambda_j)$. The $(d-1)$ dimensional parameters
$\bmu_{-1j},\bLambda_{-1j}$ stand for $\bmu_j,\bLambda_j$ but without the first component.

As a result, assuming $k$ distinct components $\btheta^*_1,\ldots,\btheta^*_k$ in $\bTheta_M$, and assuming
further that each distinct component $\btheta^*_j$ occurs $M_j$ times, we have,
\begin{equation}
E[y_{1}\mid\bTheta_M,\by_{-1}]=
\sum_{j=1}^k w^{(j)}(\by_{-1})\mu^{(j)}_{1|2,\ldots,d}
\label{eq:regression_estimation}
\end{equation}
is a weighted sum of the component regression functions $\mu^{(j)}_{1|2,\ldots,d}$, where the associated
weight $w^{(j)}(\by_{-1})$ is given by\\
\begin{center}
$w^{(j)}(\by_{-1})$ is proportional to\\
\end{center}
\begin{equation}
\frac{M_j}{M}N_{d-1}\left(\by_{-1}:\bmu^*_{-1j},{\bLambda^*}^{-1}_{-1j}\right)
\label{eq:regression_weight}
\end{equation}
and the proportionality constant is chosen such that $\sum_{j=1}^k w^{(j)}(\by_{-1})=1$.

Note that the regression function estimator developed above is structurally quite different from
that given by Muller(1996), who develop a regression estimator based on the model of EW. One clear advantage of our curve
over that of Muller(1996) is that for massive data sets the curve-fitting idea of Muller(1996) can not
be implemented due to extreme computational burden, while our curve (\ref{eq:regression_estimation}) can be easily fitted
to any data set, massive or not.

Assuming that a sample $\left\{\bTheta^{(1)}_M,\ldots,\bTheta^{(N)}_M\right\}$ is available from the posterior distribution of $\bTheta_M$
(typically by MCMC), the marginalized curve $E(y_1\mid\by_{-1})$ is estimated as
\begin{equation}
E(y_1\mid\by_{-1}) = E(E(y_1\mid\bTheta_M,\by_{-1}))\approx\frac{1}{N}\sum_{t=1}^NE(y_1\mid\bTheta^{(t)}_M,\by_{-1})
\label{eq:curve_mean}
\end{equation}
Pointwise variability of the curve is measured by
\begin{equation}
Var(y_1\mid\by_{-1})=Var(E(y_1\mid\bTheta_M,\by_{-1}))+E(Var(y_1\mid\bTheta_M,\by_{-1}))
\label{eq:curve_var}
\end{equation}
The first component of the above variance is estimated by the sample variance of $\left\{E(y_1\mid\bTheta^{(t)}_M,\by_{-1});t=1,\ldots,N\right\}$,
and the second component is estimated by the sample mean of $\left\{Var(y_1\mid\bTheta^{(t)}_M,\by_{-1});t=1,\ldots,N\right\}$.
Approximate $100(1-\tau) percent$ pointwise credible intervals of the curve are given by $E(y_1\mid\by_{-1})\pm z_{\frac{\tau}{2}}\sqrt{Var(y_1\mid\by_{-1})}$,
where $z_{\tau}$ is the $100\tau$-th quantile of a standard normal distribution.

Hence, once the MCMC realizations $\left\{\bTheta^{(1)}_M,\ldots,\bTheta^{(N)}_M\right\}$ are available, it is an easy task
to obtain a Bayesian regression curve with all summaries readily available.
In the next section we derive an extremely fast and easily implementable Gibbs sampling algorithm.

\section{Gibbs sampling implementation of the proposed model}
\label{sec:gibbs_sampling}

We assume that under $\bG_0$,\\
\begin{center}
$\left[\bLambda_j\right]$  is from\\
\begin{equation}
Wishart_d\left(\frac{s}{2},\frac{\bS}{2}\right)
\label{eq:wishart_d}
\end{equation}
$\left[\bmu_j\mid\bLambda_j\right] $ is from\\
\begin{equation}
N_d\left(\bmu_0,\psi\bLambda^{-1}_j\right)
\label{eq:normal_d}
\end{equation}
\end{center}

Hence, the joint distribution of $\btheta_j$ is given by
\begin{eqnarray}
[\Lambda_j][\bmu_j\mid\bLambda_j]&=&c|\lambda_j|^{\frac{s-d-1}{2}}\exp\left\{-tr\left(\frac{\bS\Lambda_j}{2}\right)\right\}\nonumber\\
&\times&\frac{|\bLambda_j|^{\frac{1}{2}}}{(2\pi\psi)^{\frac{d}{2}}}\exp\left\{-\frac{1}{2}(\bmu_j-\bmu_0)'\Lambda_j(\bmu_j-\bmu_0)\right\}
\label{eq:niw}
\end{eqnarray}
where
\[c=\frac{\pi^{-\frac{d(d-1)}{4}}|\frac{\bS}{2}|^{\frac{s}{2}}}{\prod_{l=1}^d\Gamma\left\{\frac{1}{2}(s+1-l)\right\}}\]

 \subsection{Representation of the mixture using allocation variables and associated full conditional distributions}
 \label{sec:fullconds}
 The distribution of $[\by_i\mid\bTheta_M]$ given by (\ref{eq:mult_normixture}) can be represented by introducing the allocation variables $z_i$, as follows:

 For $i=1,\ldots,n$ and $j=1,\ldots,M$,
 \begin{eqnarray}
 \left[\by_i\mid z_i=j,\bTheta_M\right]&=&\frac{|\bLambda_j|^{\frac{1}{2}}}{(2\pi)^{\frac{d}{2}}}\exp\left\{-\frac{1}{2}\left(\by_i-\bmu_j\right)'\bLambda_j\left(\by_i-\bmu_j\right)\right\}
 \label{eq:y_given_z}\\
 \left[Z_i=j\right]&=&\frac{1}{M}
 \label{eq:latent_z}
 \end{eqnarray}

It follows that the full conditional distribution of the allocation variables $z_i$ $(i=1,\ldots,n)$ given the rest is given by

$\left[z_i=j\mid\bY,\bTheta_M,\bZ_{-i}\right]$ is proportional to\\ 
\begin{equation}
\frac{|\bLambda_j|^{\frac{1}{2}}}{(2\pi)^{\frac{d}{2}}}\exp\left\{-\frac{1}{2}\left(\by_i-\bmu_j\right)'\bLambda_j\left(\by_i-\bmu_j\right)\right\};\hspace{2mm}j=1,\ldots,M
\label{eq:mult_full_cond_z}
\end{equation}

\subsection{Full conditionals of $\btheta_j$}
\label{subsec:mult_mucond}

Defining $n_j=\#\left\{i:z_i=j\right\}$ and $\bar \by_j=\sum_{i:z_i=j}\by_i/n_j$, we note that
the full conditional
distribution of $\btheta_j$ given the rest is given by
\begin{equation}
\left[\btheta_j\mid \bY,\bZ,\bTheta_{-jM}\right]=q_{0j}G_j(\btheta_j)+\sum_{\ell=1,\ell\neq j}^Mq_{\ell j}\delta_{\btheta_{\ell}}(\btheta_j)
\label{eq:mult_full_cond_theta}
\end{equation}
Under $G_j$ the distribution of $\btheta_j$ is given by:
\begin{eqnarray}
\left[\bLambda_j\right] & from & Wishart_d\left(\frac{s+n_j}{2},\frac{1}{2}\left\{\bS+\frac{n_j(\bar \by_j-\bmu_0)(\bar\by_j-\bmu_0)'}{n_j\psi+1}+\sum_{i:z_i=j}(\by_i-\bar \by_j)(\by_i-\bar\by_j)'\right\}\right)\nonumber\\
\label{eq:bLambda_base_posterior}\\
\left[\bmu_j\mid\bLambda_j\right] & from & N_d\left(\frac{n_j\bar \by_j\psi+\bmu_0}{n_j\psi+1},\frac{\psi\bLambda^{-1}_j}{(n_j\psi+1)}\right)\label{eq:bmu_base_posterior}
\end{eqnarray}
In (\ref{eq:mult_full_cond_theta}) $q_{0j}$ and $q_{\ell j}$ are given by the following:

$q_{0j}$ is  proportional  to $a$, where\\
\begin{eqnarray}
a&=&\alpha\frac{|\frac{\bS}{2}|^{\frac{s}{2}}}{\Gamma(\frac{s}{2})}\times\left(\frac{1}{n_j\psi+1}\right)^{\frac{d}{2}}\times\left(\frac{1}{2\pi}\right)^{\frac{n_jd}{2}}\nonumber\\
          &\times& \frac{2^{\frac{s+n_j}{2}}\prod_{l=1}^d\Gamma(\frac{s+n_j+1-l}{2})}{\left|\left\{\bS+\frac{n_j(\bar\by_j-\bmu_0)(\bar\by_j-\bmu_0)'}{n_j\psi+1}+\sum_{i:z_i=j}(\by_i-\bar\by_j)(\by_i-\bar\by_j)'\right\}\right|^{\frac{s+n_j}{2}}}
           \label{eq:mult_q0}
      \end{eqnarray}
      and,
$q_{\ell j}$ is  proportional  to\\
      \begin{eqnarray}
      \frac{|\bLambda_{\ell}|^{\frac{n_j}{2}}}{(2\pi)^{\frac{n_jd}{2}}}\exp\left[-\frac{1}{2}\left\{n_{j}(\bmu_{\ell}-\bar\by_{j})'\bLambda_{\ell}(\bmu_{\ell}-\bar\by_{j})+tr\bLambda_{\ell}\sum_{i:z_i=j}(\by_i-\bar\by_{j})(\by_i-\bar\by_{j})'\right\}\right]\nonumber\\\label{eq:mult_ql}
      \end{eqnarray}
      The proportionality constant is chosen such that $q_{0j}+\sum_{\ell=1,\ell\neq j}q_{\ell j}=1$.

It is useful to provide the intuition behind updating the allocation variables $\bZ$ and the parameters $\bTheta_M$. Given $M$ distinct values of the parameter vector $\bTheta_M$,
the allocation vector $\bZ$ clusters the $n$-dimensional observation vector $\bY$ into $M^* (\leq M)$ clusters of the form
$U_j=\{i:z_i=j\}$; $j=1,\ldots,M^*$. These can be thought of as the {\it initial clusters}, since the Dirichlet process prior acts upon $\{U_1,\ldots,U_{M^*}\}$, to
yield $k (\leq M^*)$ distinct parameter values $\btheta^*_1,\ldots,\btheta^*_k$ out of the possible $M^*$ distinct values to yield the final clustering, say, $\{V_1,\ldots,V_k\}$,
of $\{U_1,\ldots,U_M\}$, with $V_{\ell}=\cup_{j:c_j=\ell} U_j$. The clusters $V_{\ell}$ are associated
with the configuration vector $\bC$. Clearly, the clustering $\{V_1,\ldots,V_k\}$ is coarser
than $\{U_1,\ldots,U_M\}$ in the sense that the former consists of lesser number of blocks with more
elements in each block. We note a computational advantage our method over the RJMCMC algorithm of RG. Note that empty components
are naturally handled in our method; indeed, if the $j$-th component is an empty component (which can happen if the allocation
variables do not allocate any observation to the $j$-th component), then the fact $n_j=\#\left\{i:Z_i=j\right\}=0$ occurs naturally
in our model and no special care is necessary for validation of this step. But this situation requires an extra, careful,
and complicated step in the method of RG.

      \subsection{Full conditional distribution of $\alpha$}
      \label{subsec:mult_alphacond}

      This remains same as in the univariate case, which is given,
      for the prior $Gamma(a_{\alpha},b_{\alpha})$ on $\alpha$,
      given the number of distinct components $k$, and another
      continuous random variable $\eta$, by
      \begin{eqnarray}
      \left[\alpha\mid\bY,\bZ,\bTheta_M,k,\eta\right] & from & \pi_{\eta}Gamma(a_{\alpha}+k,b_{\alpha}-\log(\eta))\nonumber\\
      &+& (1-\pi_{\eta})Gamma(a_{\alpha}+k-1,b_{\alpha}-\log(\eta))
      \label{eq:full_cond_alpha}
      \end{eqnarray}
      where
      \begin{equation}
      \frac{\pi_{\eta}}{1-\pi_{\eta}}=\frac{a_{\alpha}+k-1}{m(b_{\alpha}-\log(\eta))}
      \label{eq:full_cond_eta}
      \end{equation}

      The full conditional distribution of $\eta$ given the rest is $Beta(\alpha+1,M)$, that is,
      a $Beta$ distribution with mean $(\alpha+1)/(\alpha+M+1)$.

      \subsection{Full conditional distributions of $\bmu_0$ and $\psi$}
      \label{subsec:mult_hyper_update}

      The distributions of the hyperparameters $\bmu_0$ and $\psi$ are given by:
      The distribution of $\left[\bmu_0\mid \bY,\bZ,\bTheta_M,\psi\right]$ is $d$-variate normal, with mean vector and dispersion matrix given by:
      \begin{eqnarray}
      E\left[\bmu_0\mid \bY,\bZ,\bTheta_M,\psi\right]&=&\left(\psi{\bf I}+{\bf A}\sum_{j=1}^k\bLambda^*_j\right)^{-1}\left(\psi {\bf a}+{\bf A}\sum_{j=1}^k\bLambda^*_j\bmu^*_j\right)\label{eq:e_mult_full_cond_mu0}\\
      V\left[\bmu_0\mid \bY,\bZ,\bTheta_M,\psi\right]&=& \left(\psi{\bf I}+{\bf A}\sum_{j=1}^k\bLambda^*_j\right)^{-1}{\bf A}\psi \label{eq:v_mult_full_cond_mu0}
      \end{eqnarray}
      In the above, we have denoted the identity matrix by ${\bf I}$.
      The full conditional distribution of psi is given by

$\left[\psi\mid \bY,\bZ,\bTheta_M,\bmu_0\right]$ from\\
\begin{equation}
  Gamma\left(\frac{w+K}{2},\frac{W+K}{2}\right)
  \label{eq:mult_full_cond_psi}
\end{equation}
      In (\ref{eq:mult_full_cond_psi}), $K=\sum_{j=1}^k(\bmu^*_j-\bmu_0)'\bLambda^*_j(\bmu^*_j-\bmu_0)$.

We have thus derived a simple Gibbs sampling algorithm for our mixture model with unknown number of components, which is computationally
highly suitable for massive data sets, thanks to the fixed maximum number of components.

\section{Simulation study to illustrate the performance of our curve-fitting method}
\label{sec:simstudy_curve}

We assume a bivariate normal distribution of two random variables $(y,x)$ (that is, $d=2$ in the general multivariate
methodologies developed in Sections \ref{sec:mult_extension} and \ref{sec:gibbs_sampling}), where the true regression function
of $y$ on $x$ is $x+\sin x$, a highly non-linear curve.
Pretending that the true curve is unknown, and that all we have is a sample of 1000 observations
$(x_i,y_i);i=1,\ldots,1000$, we demonstrate that our curve-fitting idea can accurately
estimate the (unknown) true curve. We obtain the data by actually simulating from the bivariate normal
distribution.

To implement our curve-fitting procedure, we need to fit the data using the Bayesian mixture
model based on Dirichlet process. Some of the prior parameters are chosen such that fast convergence to
the target posterior is ensured, and other choices (and justifications thereof) are motivated by those of EW, RG, and Bhattacharya(2008).
For example, selecting $\bmu_0$ to be the sample mean vector, and $\bS$ to be
the sample dispersion matrix indicated good mixing properties of our Gibbs sampler.
However, it is important to select the prior parameters of $\alpha$ carefully, since this can significantly
affect the probability distribution of the number of components, and hence the fit of the curve.
To select an appropriate prior for $\alpha$, we first assume that it is a constant to be determined
(by a procedure to be described below). Once it
is determined, we select the prior parameters $a_{\alpha}$ and $b_{\alpha}$ such that the mode of the prior
distribution of $\alpha$, $Gamma(a_{\alpha},b_{\alpha})$ is set equal to the determined value and the variance is
as large as possible to reflect our vagueness about the prior.

To determine the mode of the prior of $\alpha$, we fit the Bayesian curve with many fixed values of $\alpha$,
and compute the maximum absolute difference at $x_i,i=1,\ldots,1000$
between the true curve and the fitted curve $E(y\mid x)$.
We choose that value of $\alpha$ as the prior mode for
which the deviation is less than 0.4 and the fitted curve
contains most of the features of the true curve.

\begin{table}
\caption{Two-way table showing the deviations of the fitted curve from the true curve}
\label{table:table1}
\begin{center}
\begin{tabular}{|c|c|}\hline
Value of $\alpha$
 & Deviation\\
\hline
0.5 & 1.004\\
1.0 & 0.896\\
5.0 & 0.597\\
10.0 & 0.4898\\
15.0 & 0.4154\\
25.0 & 0.355\\
\hline
\end{tabular}
\end{center}
\end{table}

\begin{figure}
\includegraphics{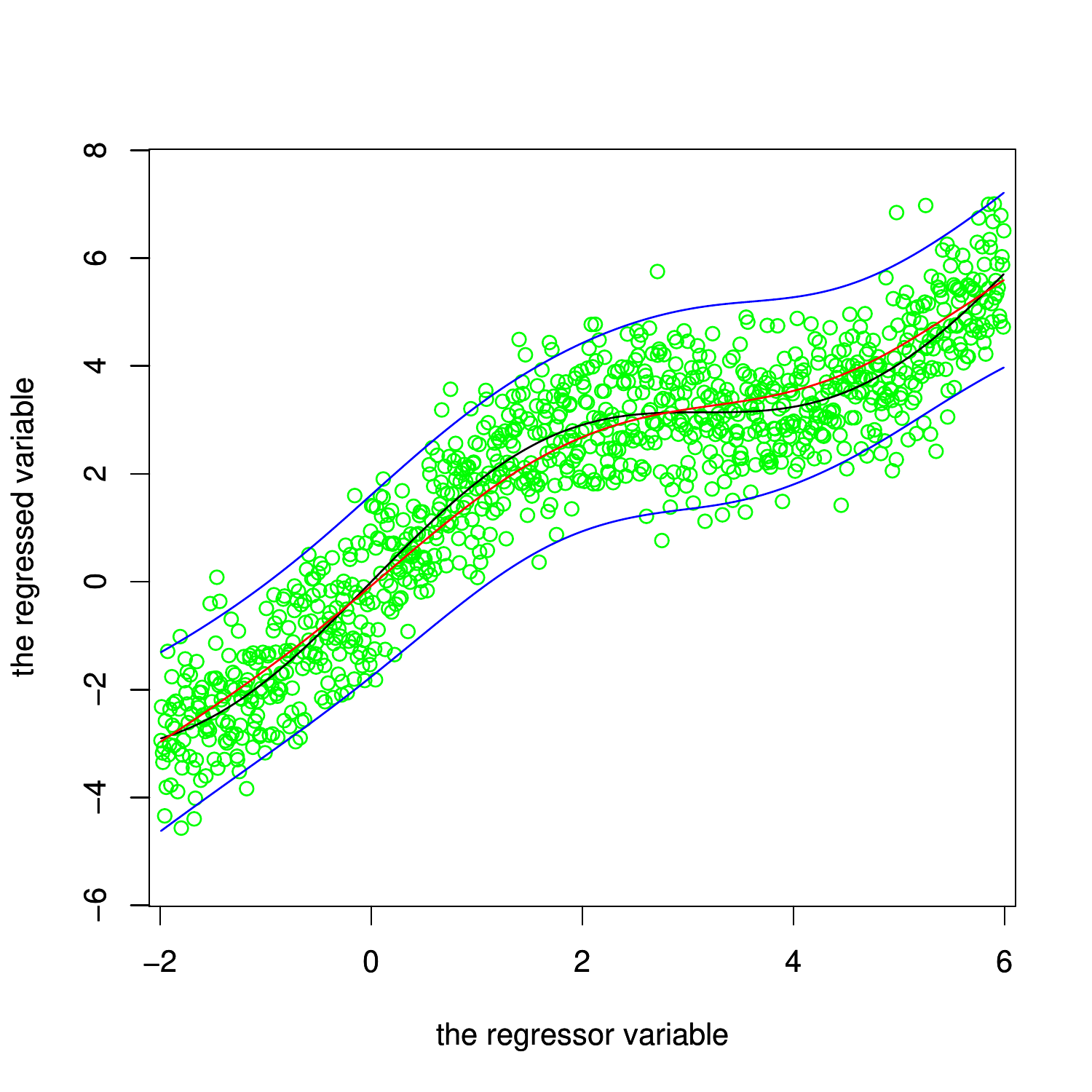}
\caption{Bayesian curve fitting: the fitted curve (green) and the true curve (yellow) associated with the simulation study.
The red curves denote pointwise 95 percent credible intervals.}
\label{fig:check(25)}
\end{figure}

Table \ref{table:table1} displays the maximum absolute deviations corresponding to
a fixed value of $\alpha$. To obtain each row of Table \ref{table:table1} we ran our Gibbs sampler
for 20000 iterations, discarding the first 5000 iterations as burn-in.
From the table we choose the value 25 as the mode of the prior distribution of $\alpha$ as the optimum
choice.

Hence, we fix the prior mode of $Gamma(a_{\alpha},b_{\alpha})$, given by $(a_{\alpha}-1)/b_{\alpha}=25$, so that
$a_{\alpha}=25b_{\alpha}+1$.
Now note that the variance of $Gamma(a_{\alpha},b_{\alpha})$ is $a_{\alpha}/b^2_{\alpha}=(25/b_{\alpha})+1/b^2_{\alpha}$. Fixing
$b_{\alpha}=0.1$ yields a considerably large variance of 350. Hence, we fix $b_{\alpha}=0.1$, which implies $a_{\alpha}=25b_{\alpha}+1=3.5$.

The associated diagram Figure \ref{fig:check(25)}, which corresponds to the derived prior choice $\alpha$ from $Gamma(3.5,0.1)$ shows that the
true regression function (yellow-coloured)
is estimated quite accurately by the fitted Bayesian semiparametric curve (green-coloured) for this choice.
Moreover, the pointwise 95 percent credible intervals (red-coloured) show that the entire true curve lies within the credible limits. This is very encouraging,
given that the true model is highly non-linear.

\section{Application of the curve-fitting procedure to the real cosmological data set}
\label{sec:real_application}

We now apply our methodologies to analyse the massive cosmological data set described in Sections \ref{sec:intro} and \ref{sec:overview}.
Recall that the data set consists of 96307 bivariate data points on logarithm of redshift ($z$) and apparent magnitude ($m$)
for Qsasars (qsasi-stellar objects) collected from SDSS data, and because of the immense number of observations, it is absolutely
impossible to implement the method of EW. The RJMCMC method of RG has been illustrated for univariate observations only, and even
in that situation the procedure is overly complicated, and, in fact, had forced an error from the authors (the corrigendum has been
provided in Richardson(1998). For bivariate observations, as in our example, the RJMCMC algorithm proposed by RG for univariate
observations is not scalable for bivariate (or multivariate) observations without serious loss of efficiency. In sharp contrast to
these popular methods, our methodologies, as we have shown, are easily and efficiently scalable to any dimensionality, and
quite importantly, is extremely fast, unlike the method of EW or other related ideas. Indeed, although it is impossible to implement
the methods of EW in our real example, our Gibbs sampling algorithm completed 20,000 iterations in just about 10 hours; considering
the enormity of the number of observations, this indicates great efficiency.
We discarded the first 5000 MCMC realizations as burn-in and stored the remaining 15,000 for inference.
Informal convergence diagnostics indicated excellent mixing properties of our algorithm. A
convergence diagnostic method suited for semiparametric mixture models has been prescribed by Bhattacharya(2008);
their method confirms excellent convergence in this example.

In this real data situation, unfortunately, the true curve is unknown, hence we can not use exactly the same
procedure as in the simulation study case to determine the prior of $\alpha$.
However, the concept of mixture models offers another interesting alternative, as detailed below.
It is well-known that as the
number of components in the mixture increases, closer is the approximation to the true curve. The price paid is the loss
of parsimony of the model, however, we can forsake parsimony only for determining the prior of $\alpha$, not for model-fitting.
So, for our purpose, we first fit a mixture model to the cosmological data with a fixed (large) number of components. Since $m=30$
was fixed as the maximum number of components in our Dirichlet process-based model, $M=30$ is a natural choice for the mixture model with fixed, but
large number components. The resulting Gibbs sampler is implemented by simulating the allocation variables from the full conditional distribution (\ref{eq:mult_full_cond_z})
but simulating the parameters $\btheta_j$ from $\bG_j$, rather than from (\ref{eq:mult_full_cond_theta}) for all iterations.

The curve thus obtained can be taken as a close approximation to the ``true'' curve. We further increased the
value of $M$ to 50 but noted no significant deviation of the resuting curve from that corresponding to $M=30$.

We then applied the prior determining procedure in the case of $\alpha$, as described
in Section \ref{sec:simstudy_curve}, given the ``approximately true" curve as obtained
by the above method.
In other words, successively fixing $\alpha$ and noting the
maximum absolute deviations of the fitted curves from the ``approximately true'' curve,
we chose the appropriate value of $\alpha$, which turned out to be 50 in this real cosmological data case.
As a consequence the prior on $\alpha$ is given by
$Gamma(26,0.5)$.

\subsection{Fitted Bayesian cosmological curve and change point analysis}
\label{subsec:Fitted}

\begin{figure}
\includegraphics[width=7in,height=7in]{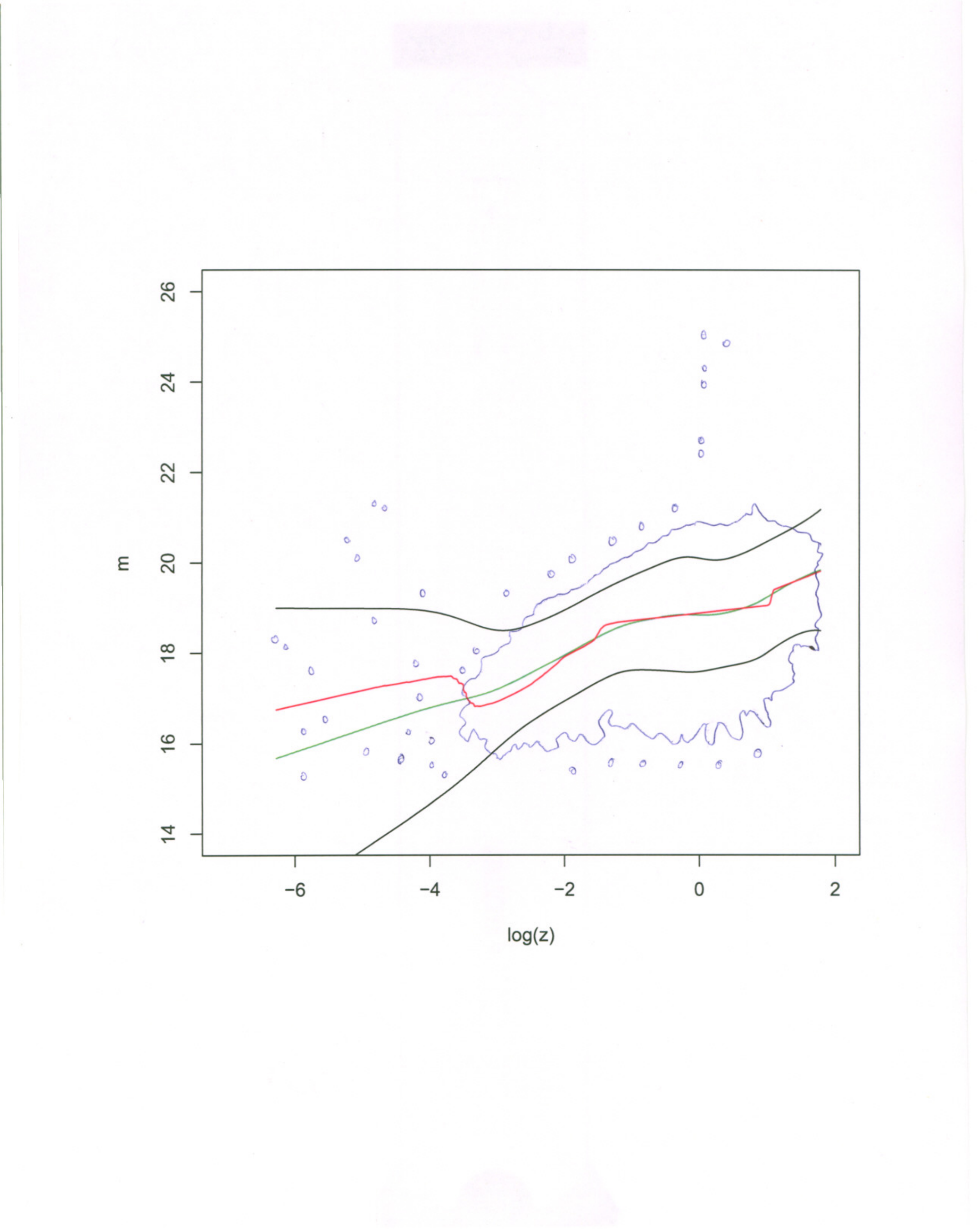}
\caption{the border of the most dense part of
the data and some of the distant lying values are shown in blue, the
fitted curve in green, the change point curve in red, and the
pointwise 95 percent credible intervals are shown in black colour.}
\label{fig:check_plot}
\end{figure}

The fitted curve and the associated pointwise 95 percent credible intervals are shown in Figure \ref{fig:check_plot} (due to difficulty detailed plot of data can not be shown); the green line
represents the estimated  Bayesian cosmological curve, and the pointwise 95 percent credible intervals are shown in black colour.
The difference in the nature of the lines in the same curve occurs due to variaton in the nature of red shift of the quasars of different ages.
The number of different such quasars is reflected in the number of distinct components of the mixture model. The different distinct components of the mixture
correspond to distributions of absolute magnitude for different ages of the clusters.
The above discussion points towards a need for detailed cluster analysis of the data set, where not only the number of clusters,
but the entire clustering is of interest to astro-physicists.

The obtained non-linear curve is linear for the first half ($z\leq -2.0$) with
intercept 18.7840 and slope 0.5136 (1.182608 with respect to logarithm with base 10, which is of interest
to astro-physicists). After that, however, non-linearity is exhibited.
But we also note that the form of non-linearity can be approximated by linear line segments indicating presence of change points. A close look will reveal presence of
four change points, but to get a solid basis of belief, we
performed a detailed change point analysis, assuming four change points.

Although a Gibbs sampling algorithm is available on the similar lines of Carlin(1992), the
algorithm is computationally expensive because of the massive number of observations. Instead, we resort to the Metropolis-Hastings
algorithm for simulating from the posterior. We omit details to save space, but remark that we achieved excellent convergence
with our Metropolis-Hastings algorithm.

Figure \ref{fig:check_plot} also shows that the curve obtained by the change point analysis, which is shown in red colour, nicely approximates
our fitted semiparametric Bayesian curve (the green curve) at all places except at the extreme lower end of the $x$-space, where there
are hardly any information about the curve.
Moreover, the entire change point curve falls within the (pointwise) 95 percent credible intervals associated with the semiparametric Bayesian
curve.

Apart from the semiparametric Bayesian curve and the change point curve, we also fitted the least squares regression line,
obtained by assuming a simple linear regression of $\log(z)$ on $M$.
This linear regression is related to Hubble's law, and the implications of the slope of this straight line will discussed later in detail.
For now we note that the least squares regression line falls well within the pointwise 95 percent credible intervals of our Dirichlet process-based
semiparameteric Bayesian curve. This shows that the linear regression, although not optimal (in the sense that normality assumption does not hold
for this data set, for example), is not ruled out by our semiparametric method.

\subsection{Estimation of the densities of the observed data and goodness of fit check}
\label{subsec:modelfit}

Note that the marginal densities of $y=m$ and $x=\log(z)$ can be estimated from our mixture model, given
the MCMC-based posterior realizations $\left\{\bTheta^{(t)}_M;t=1,\ldots,N\right\}$, for any $X=x$ and $Y=y$, as
\begin{eqnarray}
\hat f_X(x)&=&\frac{1}{N}\sum_{t=1}^N [x\mid \bTheta^{(t)}_{M}]\nonumber\\
&=& \frac{1}{M}\frac{1}{N}\sum_{t=1}^N \sum_{j=1}^{M}N(x:\mu^{(t)}_{1j},1/\lambda^{(t)}_{1j})
\label{eq:density_x}
\end{eqnarray}
and
\begin{eqnarray}
\hat f_Y(y)&=&\frac{1}{N}\sum_{t=1}^N [y\mid \bTheta^{(t)}_{M}]\nonumber\\
&=& \frac{1}{M}\frac{1}{N}\sum_{t=1}^N \sum_{j=1}^{M}N(y:\mu^{(t)}_{2j},1/\lambda^{(t)}_{2j})
\label{eq:density_y}
\end{eqnarray}
Pointwise 95 percent credible intervals can be obtained for each of the marginal densities as in the case
of Bayesian curve estimation.

\begin{figure}
\includegraphics{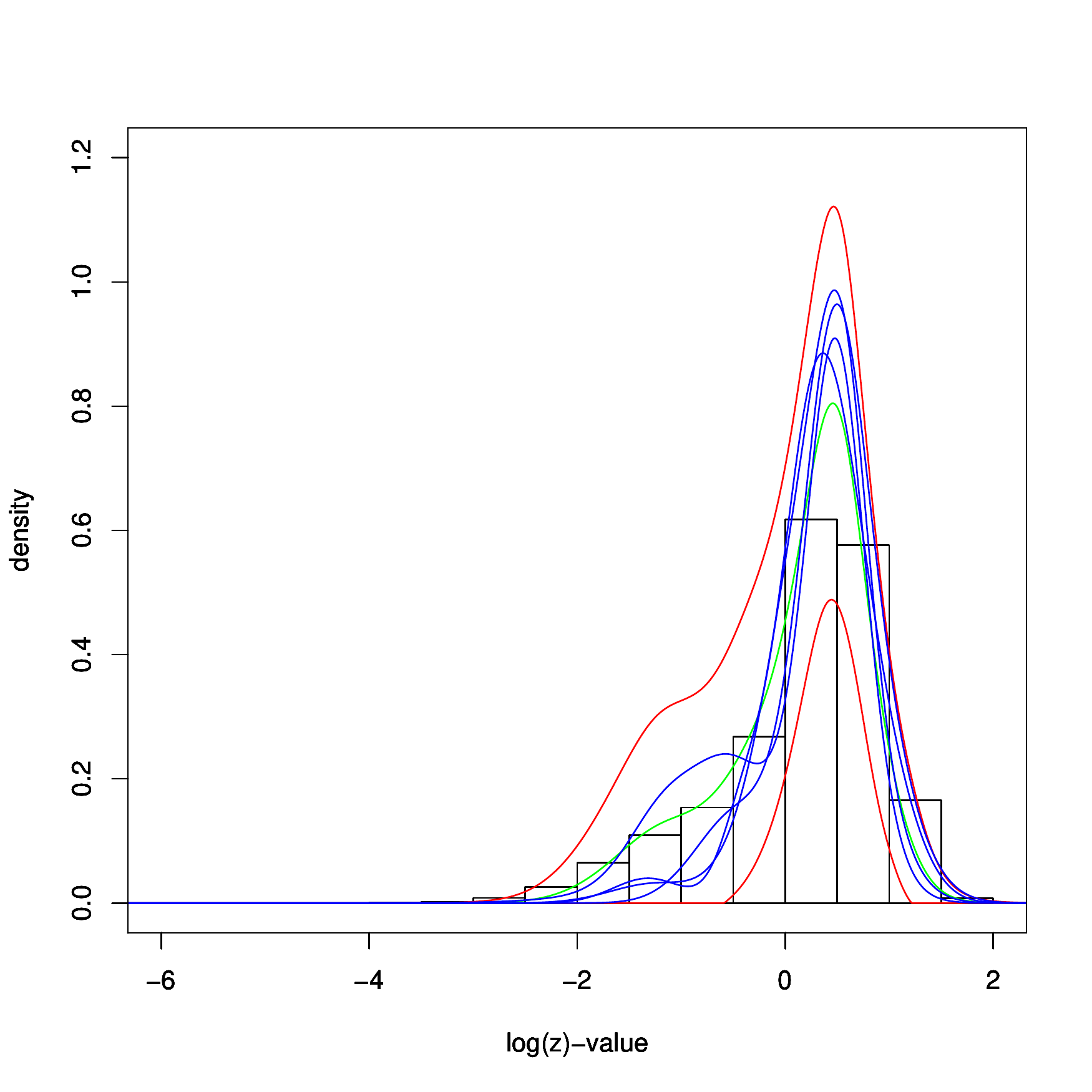}
\caption{Marginal density estimation of $\log(z)$: the red lines represent the 95 percent limits of the density, the green line stands
for the fitted density and the blue lines represent sample densities.}
\label{fig:marginalx}
\end{figure}

\begin{figure}
\includegraphics{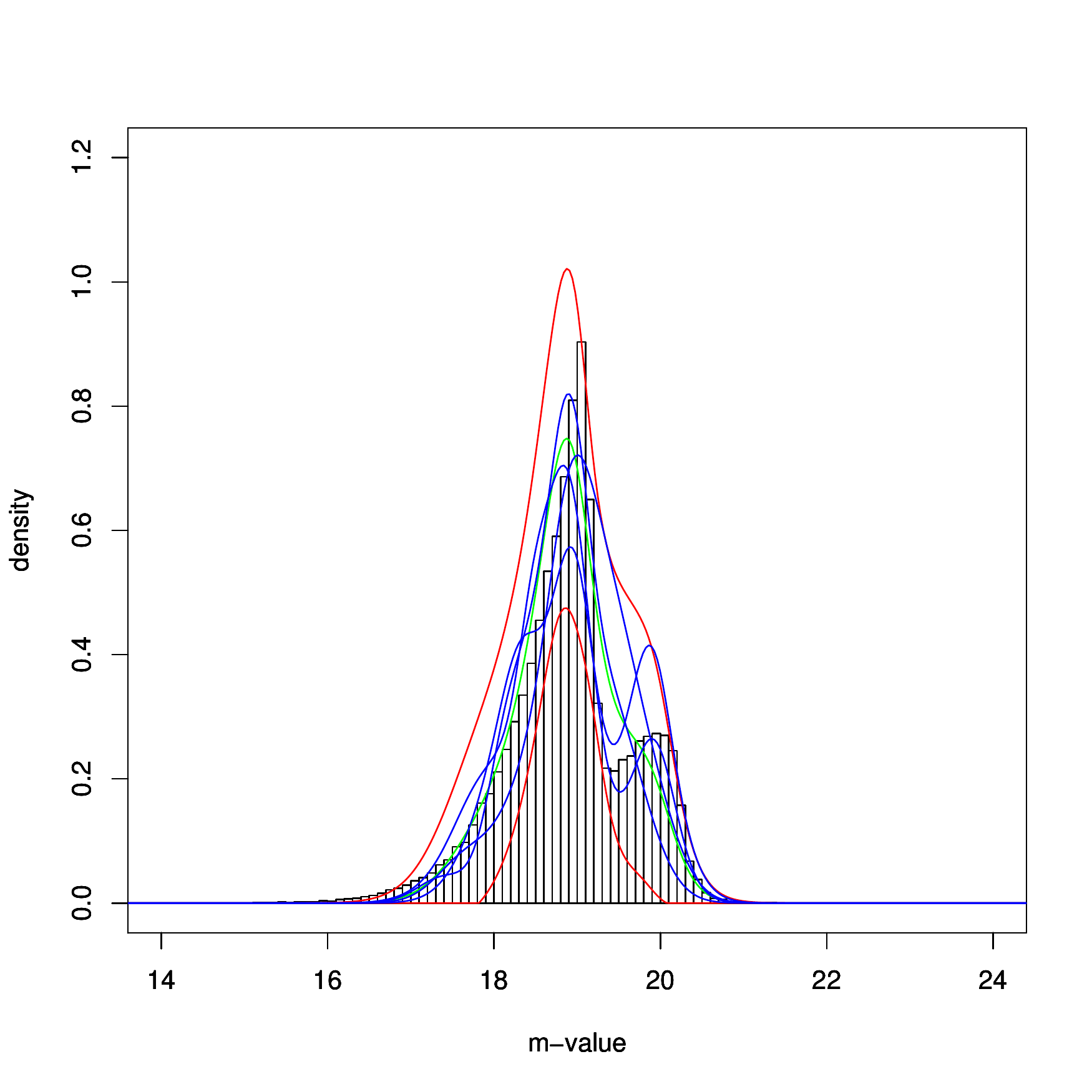}
\caption{Marginal density estimation of $m$: the red lines
represent the 95 percent limits of the density, the green line
represents the fitted density and blue lines stand for sample
densities.} \label{fig:marginaly}
\end{figure}

These Bayesian density estimates are useful for model validation purpose. In fact, these density estimates
can be compared with the observed histograms of the individual variables of the observed data. A high
degree of discrepancy between the observed histogram and the corresponding density estimate will indicate
lack of model fit.
Figures \ref{fig:marginalx} and \ref{fig:marginaly} show the observed marginal histograms, the marginal
density estimates, and the associated 95 percent credible intervals of the true density. A few sample densities are
also shown. Very clearly, the marginal density estimates fit the histogram very satisfactorily, leaving
no reason to doubt the validity of our mixture model. In fact, the histograms (if smoothed by any means),
the density estimates, and also the sample densities, all lie within their respective 95 percent credible intervals,
which is very encouraging.

\section{Bayesian posterior distribution of clustering}
\label{sec:clustering}

As discussed in Section \ref{subsec:Fitted}, it is of interest to astro-physicists to conduct a detailed study of the clusters
of the data set.
Thus, a methodology is needed which provides not only the posterior distribution of the number of clusters,
but the posterior distribution of the clusterings, using which summaries of clusterings may be obained.
We note that, our Gibbs sampling algorithm generates a clustering with varying number of clusters in each iteration (see
Section \ref{subsec:mult_mucond}; see also the Appendix, where the randomness of clusterings and the number of clusters is induced by the configuration vector $\bC$)
The fact that even if the number of clusters are same in any two iterations, the corresponding clusterings are still different, shows that it is important
to deal with the posterior distribution of clusterings rather than the posterior distribution of the number of clusters. Moreover, it is usually
of scientific interest to analyse some representative of the clusterings produced by the Gibbs sampler, as in our cosmological example.

It is to be noted that this problem is much more difficult as compared to summarization of posterior distribution of a parmeter.
In the case of a parameter the posterior distribution can be
summarized by its posterior mean or mode (analytical or sample-based).
Similarly desired credible regions can be easily calculated. But it is not possible to take means of clusterings produced.
Due to continuity of the parameters the mean will give rise to a $M$-component clustering, though all
the clusterings might consist of less than $M$ clusters. Moreover the clusterings are permutation invariant.
That is two clustering may be same except for a permutation of the components.

Construction
of credible region poses even more difficulties. Here we use a methodology introduced by Bhattacharya(2009)
to tackle such difficulties.

The methodology of Bhattacharya(2009) relies on an appropriately defined metric to compute distances between any two clusterings
of a given data set. The metric was used to compute the posterior probability distribution of clusterings, and
to provide a ``central" clustering and the associated credible regions. The authors applied
their methodology to analyse the posterior distribution of clusterings of a large
vegetation data set obtained from the Western Ghats in India, generated by the method of EW.
In this paper, we apply their cluster analysis methodology to the clusterings generated by our Bayesian
mixture model.

\subsection{Definition of central clustering}
\label{subsec:exact_central_clustering}
Guided by the definition of mode in
the case of parametric distributions, given a suitable metric $d$ to compute the distance between
any two clusterings, Bhattacharya(2009) define a clustering $C^*$ as ``central'' if, for a given small $\epsilon>0$,

\begin{equation}
  P\left(\left\{C:d(C^*,C)<\epsilon\right\}\right)=\sup_{C'} P\left(\left\{C:d(C',C)<\epsilon\right\}\right)
\label{eq:central_cluster}
\end{equation}

Thus for a sufficiently small $\epsilon>0$, the probability of an
$\epsilon$-neighbourhood of an arbitrary clustering $C$
is highest when $C = C^*$, the central clustering. The above definition holds for all positive
$\epsilon$ if the distribution of clustering is unimodal.

Otherwise the depending on $\epsilon$ we will have different local modes of clustering, from among which
the global mode is to be determined.

\subsection{Empirical Definition of Central Clustering}
\label{subsec:empirical_central_clustering}

We define that clustering $C^{(j)}$ as ``approximately central" which, for a given small $\epsilon>0$, satisfies
the following equation
\begin{equation}
C^{(j)}=\arg\max_{1\leq i\leq N}\frac{1}{N}\#\left\{C^{(k)};1\leq k\leq N:d(C^{(i)},C^{(k)})<\epsilon\right\}
\label{eq:empirical_central_cluster}
\end{equation}

The central clustering $C^{(j)}$ is easily computable, given $\epsilon>0$ and a suitable metric $d$. Also,
by the ergodic theorem, as $N\rightarrow\infty$ the empirical central clustering $C^{(j)}$ converges almost surely to the exact
central clustering $C^*$.

Given a central clustering $C^{(j)}$ one can then obtain, say, an approximate
95 percent highest posteror density credible region
as the set $\left\{C^{(k)};1\leq k\leq N:d(C^{(k)},C^{(j)})<\epsilon^*\right\}$, where $\epsilon^*$
is such that
\begin{equation}
\frac{1}{N}\#\left\{C^{(k)};1\leq k\leq N:d(C^{(k)},C^{(j)})<\epsilon^*\right\}\approx 0.95
\label{eq:hpd}
\end{equation}

In (\ref{eq:hpd}) $\epsilon^*$ must be chosen by trial and error.

\subsection{Choice of the metric $d$}
\label{subsec:metric}

Two clusterings may not be very easily comparable as the cluster number of one may totally unrelated to the cluster numbers of the other. So, one way to compare them
is to find a measure of divergence between them after permuting the arbitrary indices to make
the two clusterings as close to each other as possible.
Ghosh(2008) define the distance $d(I , II)$ between clusterings $I$ and $II$ as follows.

\begin{equation}
d(I,II)=\min [n_{00}-(n_{1j_1}+n_{2j_2}+\ldots +n_{kj_k})]/n_{00}
\label{eq:metric}
\end{equation}
over all permutations $(j_1,j_2,\ldots,j_k)$ of $(1,2,\ldots,k)$, where $k$ denotes the number of
clusters, $n_{ij}$ is the number of units belonging to the $i$-th cluster of $I$ and $j$-th cluster of $II$,
and $n_{00}=\sum\sum n_{ij}$ is the total number of units.
For justification of the above idea, and for the proof that (\ref{eq:metric}) satisfies the properties of a metric, see Ghosh(2008).

However, computation of the above metric (\ref{eq:metric}) requires the
minima over all possible permutations of the clusters.
If the number of clusters under consideration is large this leads to enormous computational burden. For MCMC iterations,
one needs to compute the metric for a large number of clusterings
(one for each iteration),
and since each iteration may yield quite a large number of clusters, the calculation quickly becomes infeasible.
To overcome this Bhattacharya(09) propose an approximation to (\ref{eq:metric}) as
\begin{equation}
\hat d(I,II)=\max\left\{\tilde d(I,II),\tilde d(II,I)\right\}
\label{eq:approx_final}
\end{equation}
where
\begin{eqnarray}
\tilde d(I,II)&=&\left\{n_{00}-\sum_{i=1}^k\max_{1\leq j\leq k}n_{ij}\right\}\bigg/n_{00}\label{eq:approx1}\\
&=&1-\frac{\sum_{i=1}^k\max_{1\leq j\leq k}n_{ij}}{n_{00}}\label{eq:approx2}
\end{eqnarray}

Very clearly, no computational labour is required to compute $\hat d(I,II)$. Very importantly, Bhattacharya(09) demonstrate
that $\hat d$ provides very accurate approximations to the original metric $d$.
Moreover, it is easy to see that $\hat d$ satisfies first three properties of a metric. The fourth property can be seen to be valid when the clusterings are independent.
But no counter example has been so far come across.
So Bhattacharya(09) conjecture that $\hat d$ is a metric.
As a result, for our analysis we will always use $\hat d$ instead of $d$.

\section{Application of the clustering idea to the cosmology data set}
\label{sec:clustering_application}

On application of the central clustering ideas, we observe that for different range of values of $\epsilon>0$ we have different
central clusterings, clearly indicating multimodality of the posterior distribution of clusterings.
For $0<\epsilon<0.05$ the central clustering is 1138-th clustering after considering burn-in. For $0.05<\epsilon<0.1$ it is 4341-th; for  $0.1<\epsilon<0.3$
it is 4849-th; for  $0.3<\epsilon<0.5$
the number is 570-th clustering after considering burn-in etc.
Following the technique Bhattacharya(09) applied for obtaining the global central clustering, we
obtain the clustering corresponding to iteration number 1137 as the global central clustering.
The radius of 95 percent credible region of the global mode is 0.35, which is reasonably low.

We note that the central clustering in our case consists of 29 clusters. This is quite reasonable, given that
there are more than 96,000 observations. Moreover, we note that although there are 29 clusters, many are effectively
the same cluster, thanks to the small Euclidean distances between them.
This reduction of the effective number of clusters finds reasons more than statistical within the astro-physics paradigm.
Indeed, astro-physicists (for example, Roy(2007)) have tried to split the data into 2 characteristics only, namely Radio loud and Radio quiet.

\begin{figure}
\includegraphics[width=7in,height=7in]{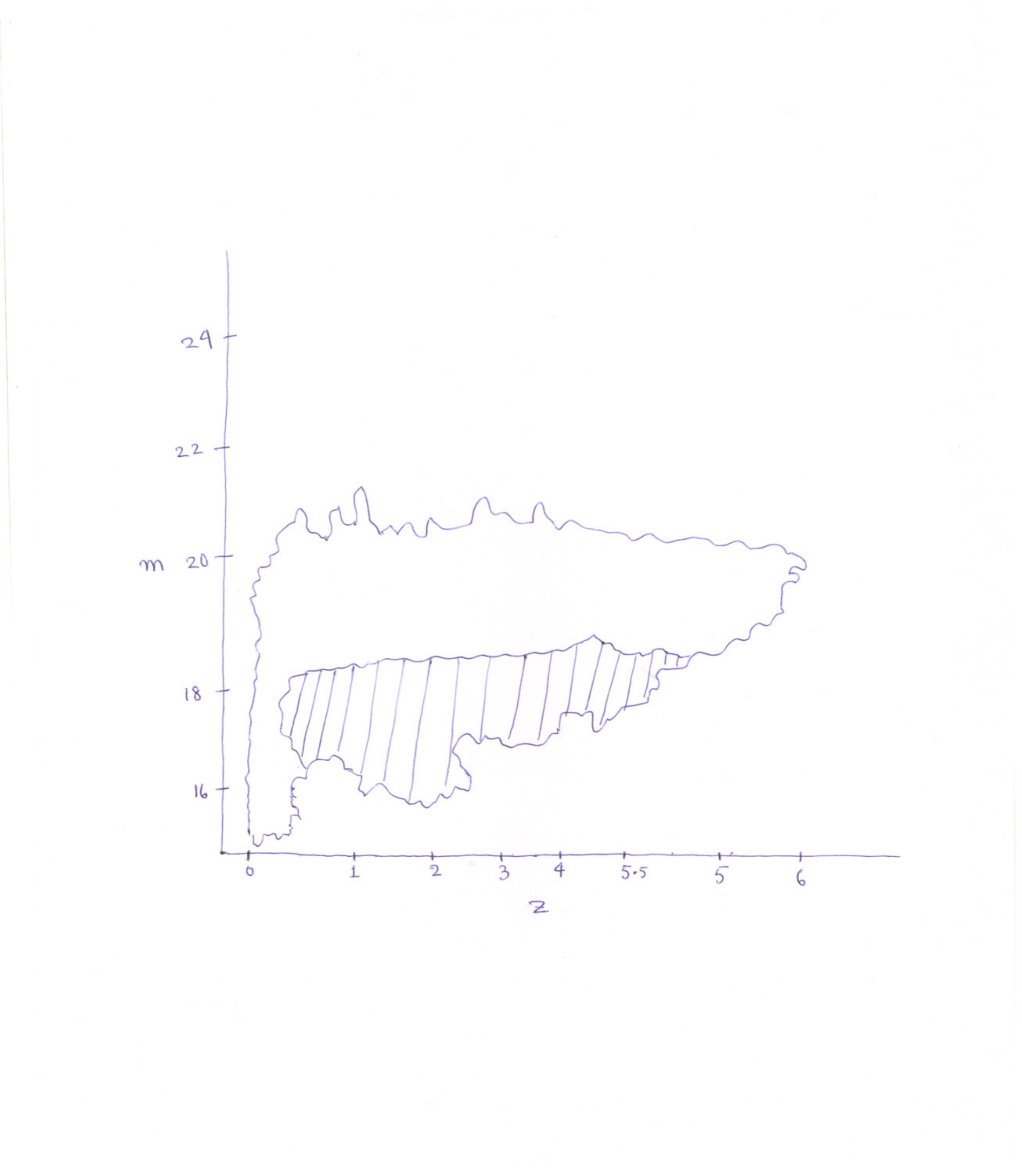}
\caption{Merged central clustering with 2 clusters: the
differently marked parts indicate two different clusters.}
\label{fig:2clusters}
\end{figure}

Driven by the above observations and discussions, we merge those clusters with Euclidean distances less than a prefixed limit.
Table \ref{table:table3} shows how the number of clusters change if the prefixed limit is changed.
\begin{table}
\caption{Table showing the variation in the number clusters with change in the prefixed limit}
\label{table:table3}
\begin{center}
\begin{tabular}{|c|c|}\hline
Value of prefixed limit
 & Number of clusters after merger\\
\hline
0.05 & 23\\
0.1 & 21\\
0.3 & 10\\
0.5 & 9\\
0.65 & 5\\
0.7 &  4\\
0.9 &  2\\
\hline
\end{tabular}
\end{center}
\end{table}
The merged central clustering
consisting of two components only (which corresponds to the prefixed limit being 0.9) is shown
in Figure \ref{fig:2clusters}. This is provided
to make our analysis comparable to the clustering done by astro-physicists on the basis of
Radio loud and Radio quiet.

We have repeated the same analysis taking the maximum number of components,
$M=50$. No notable difference between the
results of the two analyses were observed.
Indeed, even with $M=30$, the posterior probability of 30 components turned out
to be negligible (about $0.001$).

\section{Possible implications of the results of statistical analysis}
\label{sec:implications}

Our statistical analysis of the SDSS data has a number of implications that may give
answer to many interesting questions from view point of quasar astronomy and cosmological models.
\begin{itemize}

\item  The curve is linear for the
small values of $z$ and  becomes nonlinear for high values. It is to be noted that for low redshift $z$ the curve is linear
with gradient of 0.5136 (1.182608 when logarithm of $z$ is taken in log base 10 while fitting the curve) and intercept 18.7840.
In case of standard cosmological model, the
gradient of the Hubble line is supposed to lie  between $4.8$ and  $5.3$
Efron(92). One of the present authors SR Roy(07)
analysed the quasar data using non-parametric methods developed by Efron(92)
both for Veron Cetty as well as for SDSS DR-3 data and found also linearity for small $z$ and non-linearity for high $z$.

\item The whole curve can be approximated by five line segments with
four change points.

\item

We performed a detail cluster analysis that gives
rise to possible number of clusters as 29. But from observing
values of the components it is evident that the clusters can be
merged further to fewer number of clusters. The degree of
reduction depends upon the prefixed threshold for merging the
clusters.

\item The merged clusters can be compared with the clusterings
observed by astronomers Porciani(2006)
for example, redshift dependent
clusters or luminosity dependent clusters. This will be discussed elaborately in future communications.

\item There exists two broad class of quasars like radio-loud and radio quiet quasars. The
environments around these clusters of quasars are different. The width and other characteristics of the emission or
absorption lines from these quasars will be affected depending on
the nature of the environments.

\item In our framework, we have merged
the clusters into two broad clusterings under certain thresholds.
The characteristics of these clusterings need to be investigated
in details so as to compare with the radio loud or radio quiet
quasars which will be done in subsequent publications.
\end{itemize}

The non-linearity of the curve may
be due to several factors like evolution of the quasars,
acceleration or deceleration of the universe. Our findings will
shed new light not only on the validity of Hubble law but also
help us to estimate the acceleration/deceleration parameters.
These issues are very much important from the point of view of
cosmological debates and will be considered in details in
subsequent publications.

\section*{Appendix}

\begin{appendix}

\section{Reparameterization using configuration indicators}
\label{sec:configuration}
Let $k_{j}$ denote the number of distinct values in $\bTheta_{-jM}$, and let $\btheta^{j^*}_{\ell}$; $\ell=1,\ldots,k_{j}$ denote
the distinct values. Also suppose that $\btheta^{j^*}_{\ell}$ occurs $M_{\ell j}$ times.
Then a reparameterization of our model parameters can be devised as follows.

As in Muller(96), we introduce the configuration vector $\bC=\left\{c_1,\ldots, c_M\right\}$, where $c_j=\ell$
if and only if $\btheta_j=\btheta^*_{\ell}; j,\ell=1,\ldots,M$. The configuration vector $\bC$ thus provides
a reparameterization of the original parameters, the latter being reparameterized into distinct components
and the associated configuration vector. Using this reparameterized version one can can avoid simulation
of all the parameters corresponding to all $m$ components. In fact, once a configuration is simulated, only the distinct
parameters may be simulated. Moreover, the corresponding Gibbs sampler may have superior convergence properties (see MacEachern(94)).

\subsection{Full conditional distributions of the distinct values of $\bTheta_m$}
\label{subsec:mult_distinct_theta_fullcond}

The conditional posterior distribution of $\btheta^*_{\ell}$ is given by\\
$\left[\btheta^*_{\ell}\mid \bY,\bZ,\bC\right]$ from\\
\begin{equation} Wishart_d\left(\bLambda^*_{\ell}:s^*_{\ell},\bS^*_{\ell}\right)\times N_d\left(\bmu^*_{\ell}:\bmu^*_0,\psi^*_{\ell}{\bLambda^*_{\ell}}^{-1}\right)
\label{eq:mult_theta_star_given_C}
\end{equation}
In the above, $n^*_{\ell}=\sum_{j:c_j=\ell}n_j$, $\bar \by^*_{\ell}=\sum_{j:c_j=\ell}n_j\bar \by_j\Big/\sum_{j:s_j=\ell}n_j$,
$s^*_{\ell}=\frac{n^*_{\ell}+s}{2}$, $\bS^*_{\ell}=\frac{1}{2}\left\{\bS+\frac{n^*_{\ell}(\bmu_0-\bar \by^*_{\ell})(\bmu_0-\bar \by^*_{\ell})'}{\psi n^*_{\ell}+1}
+\sum_{j:c_j=\ell}n_j(\bar \by_j-\bar \by^*_{\ell})(\bar \by_j-\bar \by^*_{\ell})'+\sum_{j:c_j=\ell}\sum_{i:z_i=j}(\by_i-\bar\by_j)(\by_i-\bar\by_j)'\right\}$, and
$\psi^*_{\ell}=\psi\big / \left(\psi n^*_{\ell}+1\right)$. It is to be noted that the $\btheta^*_{\ell}$ are conditionally independent.

      \subsection{Full conditional distributions of the configuration indicators $c_j$}
      \label{subsec:mult_config_fullcond}

      The conditional distributions of $c_j$ are given, in the multivariate case, by\\
$[c_j=\ell\mid \bY,\bZ,\bC_{-j},\bTheta_M]$ is proportional to\\ 
      \begin{equation}
       \left\{\begin{array}{c}q^*_{\ell j}\hspace{2mm}\mbox{if}\hspace{2mm}\ell=1,\ldots,k_j\\q_{0j}\hspace{2mm}\mbox{if}\hspace{2mm}\ell=k_j+1\end{array}\right.
      \label{eq:mult_config_fullcond}
      \end{equation}
      where $q_{0j}$ is the expression given by (\ref{eq:mult_q0}), and\\
$q^*_{\ell j}$ is proportional to\\
      \begin{eqnarray}
       M_{\ell j}\frac{|\bLambda^*_{\ell}|^{\frac{n_j}{2}}}{(2\pi)^{\frac{n_jd}{2}}}\exp\left[-\frac{1}{2}\left\{n_{j}(\bmu^*_{\ell}-\bar\by_{j})'\bLambda^*_{\ell}(\bmu^*_{\ell}-\bar\by_{j})+tr\bLambda^*_{\ell}\sum_{i:z_i=j}(\by_i-\bar\by_{j})(\by_i-\bar\by_{j})'\right\}\right]\nonumber\\\label{eq:mult_ql_star}
      \end{eqnarray}
      Note that it is possible to replace $q_{0j}$ in (\ref{eq:mult_q0}) with
      \begin{equation}
      \alpha\frac{|\bLambda^*_{j}|^{\frac{n_j}{2}}}{(2\pi)^{\frac{n_jd}{2}}}\exp\left[-\frac{1}{2}\left\{n_{j}(\bmu^*_{j}-\bar\by_{j})'\bLambda^*_{j}(\bmu^*_{j}-\bar\by_{j})+tr\bLambda^*_{j}\sum_{i:z_i=j}(\by_i-\bar\by_{j})(\by_i-\bar\by_{j})'\right\}\right]\nonumber\\
      \label{eq:mult_nonconjugate_form}
      \end{equation}
      where $\btheta^*_j=(\bmu^*_j,\bLambda^*_j) from \bG_0$.
      The latter formulation is most appropriate when $\bG_0$ is not conjugate to the likelihood, which may preclude integration
      of (\ref{eq:mult_nonconjugate_form}) with respect to $\bG_0$, making the explicit form of $q_{0j}$ intractable.
      In our case, we can also integrate $q^*_{\ell j}$ with
      respect to the conditional posterior distribution of $\btheta^{j^*}_{\ell}$ given by (\ref{eq:mult_theta_star_given_C}) to obtain
      \begin{equation}
      q^{**}_{\ell j}=M_{\ell j}\left(\frac{1}{\pi}\right)^{\frac{n_jd}{2}} \left(\frac{\psi n^*_{\ell}+1}{\psi n^*_{\ell}+\psi n_j+1}\right)^{\frac{d}{2}}\times\prod_{l=1}^d\frac{\Gamma\left(\frac{n^*_{\ell}+n_j+s+1-l}{2}\right)}{\Gamma\left(\frac{n^*_{\ell}+s+l-1}{2}\right)}\times\frac{S^{(1)}_{\ell}}{S^{(2)}_{\ell}}
      \label{eq:mult_integrated_ql}
      \end{equation}
      In (\ref{eq:mult_integrated_ql})
      \begin{eqnarray}
      S^{(1)}_{\ell}&=&\Big|\bS+\frac{n^*_{\ell}(\bmu_0-\bar\by_{\ell})(\bmu_0-\bar\by_{\ell})'}{\psi n^*_{\ell}+1}+\sum_{j:s_j=\ell}n_j(\bar\by_j-\bar\by^*_{\ell})(\bar\by_j-\bar\by^*_{\ell})'\nonumber\\
      &+&\sum_{j:s_j=\ell}\sum_{i:z_i=j}(\by_i-\bar\by_j)(\by_i-\bar\by_j)'\Big|^{\frac{n^*_{\ell}+s}{2}}
      \label{eq:mult_s1_l}
      \end{eqnarray}
      and
      \begin{eqnarray}
      S^{(2)}_{\ell}&=&\Bigg|\bS+\frac{n^*_{\ell}(\bmu_0-\bar\by_{\ell})(\bmu_0-\bar\by_{\ell})'}{\psi n^*_{\ell}+1}+\sum_{i:z_i=j}(\by_i-\bar\by_j)(\by_i-\bar\by_j)'\nonumber\\
      &+& \sum_{j:s_j=\ell}n_j(\bar\by_j-\bar\by^*_{\ell})(\bar\by_j-\bar\by^*_{\ell})'+\sum_{j:s_j=\ell}\sum_{i:z_i=j}(\by_i-\bar\by_j)(\by_i-\bar\by_j)'\nonumber\\
      &+&\frac{n_j[(\bar\by_j-\bmu_0)+\psi n^*_{\ell}(\bar\by_j-\bar\by^*_{\ell})][(\bar\by_j-\bmu_0)+\psi n^*_{\ell}(\bar\by_j-\bar\by^*_{\ell})]'}{(\psi n^*_{\ell}+1)(\psi n^*_{\ell}+\psi n_j+1)} \Bigg|^{\frac{n^*_{\ell}+n_j+s}{2}}
      \label{eq:mult_s2_l}
      \end{eqnarray}

\end{appendix}

\end{document}